\documentclass[a4paper,11pt]{article}
\usepackage{pos}
\usepackage{caption}
\usepackage{subcaption}
\title{Study of heavy-flavor jets in a transport approach}

\author*[a,b]{Weiyao Ke}
\author[b,c]{Xin-Nian Wang}
\author[d]{Wenkai Fan}
\author[d]{Steffen Bass}

\affiliation[a]{Department of Physics, University of California, Berkeley\\
366 LeConte Hall, Berkeley CA 94720, United States}

\affiliation[b]{Nuclear Science Division, MS 70R0309\\
Lawrence-Berkeley National Laboratory\\
1 Cyclotron Rd, Berkeley CA 94720, United States}

\affiliation[c]{Key Laboratory of Quark and Lepton Physics (MOE) and Institute of Particle Physics, \\
Central China Normal University, Wuhan 430079, China}

\affiliation[d]{Department of Physics, Duke University,\\
Physics Bldg. Science Dr., Durham NC 27707, United States}

\emailAdd{WeiyaoKe@lbl.gov}
\emailAdd{xnwang@lbl.gov}
\emailAdd{wf39@duke.edu}
\emailAdd{steffen.bass@duke.edu}

\abstract{Measurements at the RHIC and the LHC have observed flavor dependence of single-hadron suppression, which reveal the role played by quark masses in the parton interactions with the quark-gluon plasma (QGP) medium. In this study, we explore the manifestation of quark mass effect and flavor dependence in jet observables. 
We approach this study using the LIDO transport model. Both elastic and medium-induced radiative processes are implemented for hard parton evolution in the medium. 
To guarantee energy-momentum conservation in the model for the study of full jet observables, we also include a component that mimics the energy-momentum transported by medium excitation.
We first predict the heavy-jet (B-jet, D-jet) and inclusive-jet nuclear modification factor $R_{AA}$ in central nuclear collisions at both the RHIC and the LHC beam energies. 
We observe a flavor-dependent jet suppression as a function of jet transverse momentum, which can be tested by future precision measurements of heavy jets.
We further investigate a novel observable that considers the angular correlation between two hard objects: a D-meson and a jet, which provides model constraints in addition to those imposed by inclusive measurements. }

\FullConference{%
  HardProbes2020\\
  1-6 June 2020\\
  Austin, Texas}

\begin{document}
\maketitle

\section{Introduction}
Heavy-flavor particles produced in relativistic heavy-ion collisions are flavor-tagged hard probes of the deconfined quark-gluon plasma (QGP); it also reveals the mass effects in single-parton evolution in the QGP. 
Its large mass suppresses the gluon radiation at low transverse momentum ($p_T\lesssim M$) where interactions between heavy quarks and medium are dominated by elastic collisions \cite{Moore:2004tg}. 
At high-$p_T$ ($\gg M$) gluon radiation dominates heavy-flavor energy loss, yet the its finite mass modulates radiation intensity, causing its radiative energy loss to be smaller than that of a light quark. 
The impact of the radiative process on heavy-flavor suppression has been studied in different approaches, including the transport equation and the modified QCD evolution equation \cite{Cao:2013ita,Kang:2016ofv}.

In this proceeding, we study the flavor dependence of jet observables.
Like the inclusive hadron suppression, jet modification is sensitive to parton interaction in the medium. However, since a jet is a scale-dependent many-body system, its modification also depends on the partonic configuration of the jets at the scale that they interact with the medium.
Therefore,  studying heavy-flavor jets in nuclear collisions is not only another way to approach the mass-dependent partonic interaction, but also a chance to test the medium effects on different jet samples due to the additional heavy-flavor triggering.
Finally, one can study a hard-hard type of correlations between the heavy-flavor particles and jets. Compared to inclusive measurements, this novel information helps to further constrain the transport modeling.

\section{The model}
The initial hard process is generated by Pythia8 and the parton shower evolution is evolved down to scale $Q_0$. Then, the parton shower in-medium evolution is described by the LIDO transport model developed in \cite{Ke:2018jem,Ke:2018tsh}. The $Q_0$ scale is chosen to be the typical momentum broadening of a parton in an interested $p_T$ range\footnote{We used $Q_0=0.8$ GeV at the RHIC energy where the jet $p_T$ is below 30 GeV, and used $Q_0=1.3$ GeV at the LHC energy where measurements are made for up to 100 GeV and 1 TeV jets. }.
Partons (both light and heavy) interact with the medium through both elastic and induced radiation processes.
In particular, the Landau-Pomeranchuk-Migdal (LPM) effect is implemented as described in \cite{Ke:2018tsh}, which is tested to demonstrate a good accuracy in the deep-LPM region compared to theoretical results \cite{Arnold:2008zu}.
We use the leading order running coupling constant  $\alpha_s(\max\{Q, \mu_{\min}\})$ with a minimum medium scale cut-off $\mu_{\min} \propto T$ to describe the coupling in a hot medium with temperature $T$.

For heavy quarks, the exact mass effect in the radiation rate is complicated, and what has been implemented in LIDO is the so-called dead-cone approximation\footnote{The range of validity of the dead-cone approximation was studied in \cite{Armesto:2003jh}}.
In this approximation, the gluon radiation rate is suppressed relative to that for the light quark by a ``dead-cone'' factor $\left[1+\theta_D^2/\theta^2\right]^{-2}$, where $\theta$ is the angle of the emitted gluon relative to the direction of the quark, and $\theta_D=M/E$ is the characteristic dead-cone size defined by the mass and energy of the heavy quark.

To study jet, we need to guarantee that the model conserves energy-momentum when integrated over full particle acceptance $(\eta, \phi)$.
We used a simple hydrodynamic-motivated ansatz of jet energy loss induced medium excitation. The energy and momentum density perturbation that propagates in direction $\hat{\bf k}$, induced by the parton energy loss ($\Delta p^\mu$) to soft particles ($E<4T$), is given by
\begin{eqnarray}
\frac{d\delta e}{d\hat{\bf k}} =\frac{1}{4\pi} \left(\Delta p^0 +c_s \hat{\bf k}\cdot \Delta \mathbf{p}\right), \quad \frac{d\delta {\bf p}}{d\hat{\bf k}} =\frac{3\hat{\bf k}}{4\pi} \left(\Delta p^0/c_s + \hat{\bf k}\cdot \Delta \mathbf{p}\right)
\end{eqnarray}
where $c_s^2=1/3$ is the conformal speed-of-sound for this simple model.
Medium excitation affects jet definition by correcting the transverse momentum density $dp_T/d\eta d\phi$, which is obtained by assuming the distribution function at freezeout to consist of massless particles with an average radial flow profile ${\bf v_r}=0.6\hat{\bf k}$ in central nuclear collisions.

\section{Results}

\begin{figure}
\begin{subfigure}[t]{0.48\textwidth}
         \centering
        \includegraphics[height=.55\textwidth]{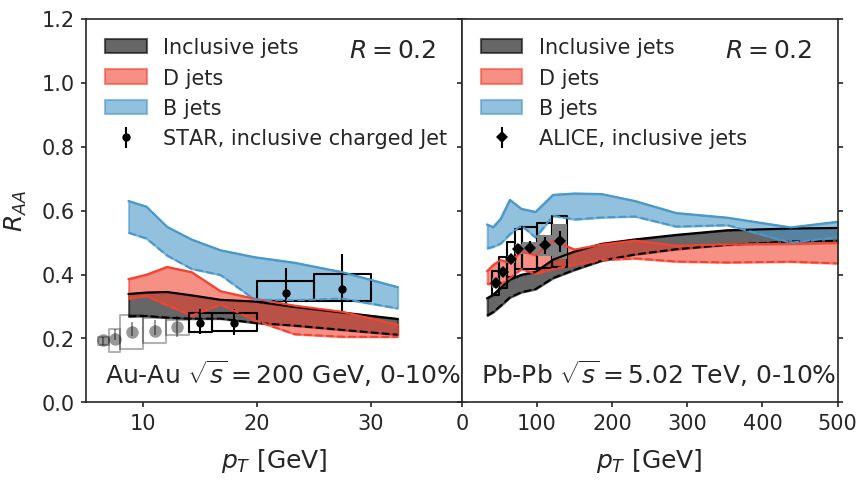}
         \caption{The left plot shows LIDO full jet calculation compared to STAR measurement of inclusive charged jets\cite{Adam:2020wen}. The right plot shows LIDO full jet calculation compared to ALICE measurement of inclusive jets \cite{Acharya:2019jyg}.}
        \label{fig:heavy-jet}
\end{subfigure}
\hfill
\begin{subfigure}[t]{0.48\textwidth}
         \centering
        \includegraphics[height=.55\textwidth]{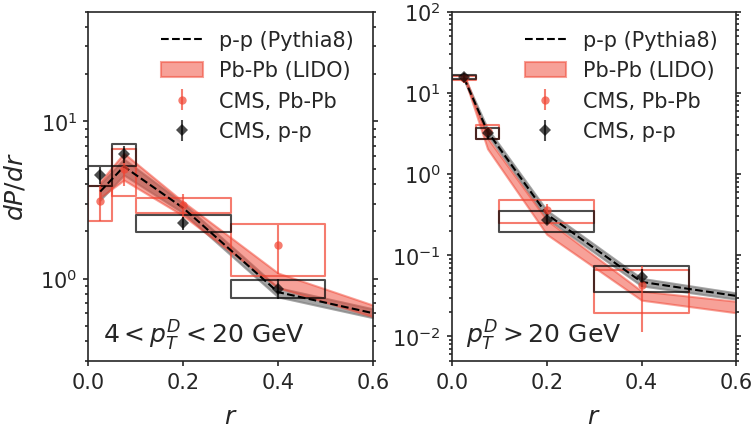}
         \caption{LIDO calculations of D-meson-jet angular correlations compared to CMS measurements \cite{Sirunyan:2019dow}. Jets have $p_T>60$  GeV and $|\eta|<2.1$. Left: $4<p_T^D<20$ GeV; right: $p_T^D>20$ GeV.
         }
        \label{fig:D-jet-corr}
\end{subfigure}
\caption{LIDO calculations of flavor-dependent jet suppression and heavy-flavor-jet angular correlations.}
\end{figure}

Figure \ref{fig:heavy-jet} shows the calculation of inclusive jet and heavy-flavor jet at both RHIC and the LHC beam energies. The bands correspond to a variation of in-medium coupling strength parameter $\mu_{\min}$ from $1.5\pi T$ to $2\pi T$, and they are compared to inclusive jet and charged-jet measurements by the ALICE and the STAR collaboration\footnote{Note that STAR inclusive jet measurement in nuclear collisions used a leading particle trigger, which biases the low-$p_T$ jet. The estimated unbiased data points are displayed in back color, while biased data points are labeled by gray.}.
At both RHIC and the LHC energies, the calculations demonstrate a flavor dependence of jet quenching. Especially $R_{AA}^{\textrm{B-jets}}$ are clearly separated from $R_{AA}^{\textrm{D-jets}}$ and that for inclusive jets, except for $p_T>300$ GeV jets at the LHC where the ordering of $R_{AA}$ changes. Such flavor dependence is not entirely the consequence of the mass effect in single parton interactions, as we observe that qualitatively similar flavor dependence persists if we remove the dead-cone factor for gluon radiations from heavy quarks. The remaining differences can be traced to a heavy-flavor trigger that selects a different jet sample than the inclusive jet sample, and such difference can be quantified by measuring the internal structure of heavy-jets compared to inclusive jets.

In addition to  jet suppression, we also study a recent CMS measurement of $D^0$-jet angular correlation\cite{Sirunyan:2019dow}. This correlation is sensitive to both the diffusion of D-meson momentum and the drifting of the jet axis in the medium. 
The data suggest a widening of $D^0$-jet angular correlation for a lower D-meson momentum range $4<p_T<20$ GeV in Pb-Pb compared to p-p due to medium effects; while the distributions for high-$p_T$ D mesons are consistent in nuclear and proton-proton collisions. This observable has been studied in another transport-based model \cite{Wang:2019xey}. Our calculation, however, predicts an angular distribution in Pb-Pb that is narrower than expected (using references generated by Pythia8) for both low-$p_T$ and high-$p_T$ D meson selections.
Because the direction of a high-$p_T$ (>20 GeV) D-meson is not strongly altered by medium effects, the current discrepancy from experiments (despite the uncertainty) suggests that there is an insufficient drifting of the jet axis from medium effect in the calculation.

\section{Summary}
In summary, using a transport model that implements both elastic and radiative processes of hard partons in the QGP medium and a parametrized medium excitation ansatz, we predict  the flavor-dependent jet suppression in central nuclear collisions at both the RHIC and the LHC energies.
We also study the medium modified $D^0$-jet angular correlation, which is sensitive to the diffusion of the jet axis relative to the diffusion of heavy flavor particles in the QGP. Compared to data, our present calculation underestimates the width of the correlation function in nuclear collisions. Further study is needed to  better understand the diffusion of the jet axis in our model.

\section*{Acknowledgement}
WK is supported by the UCB-CCNU Collaboration Grant. XW is in part supported by NSFC under Grant Nos. 11935007, 11221504, and 11890714, by DOE under  No. DE-AC02-05CH11231, and by NSF  under No. ACI-1550228 within the JETSCAPE Collaboration. WF and SAB are supported by DOE under Grant  No. DE-FG02-05ER41367, and by NSF under No. ACI-1550300.

\end{document}